\documentclass{aa}
\usepackage{graphicx}
\begin{document}
\title{Magnetohydrodynamic waves in the pulsar magnetosphere} 
\author{V.~Urpin\inst{1,2}}
\institute{$^{1)}$ INAF, Osservatorio Astrofisico di Catania, Via
           S.Sofia 78, 95123 Catania, Italy  \\
           $^{2)}$ A.F.Ioffe Institute of Physics and Technology and
           Isaac Newton Institute of Chile, Branch in St. Petersburg,
           194021 St. Petersburg, Russia
}
\date{\today}

\abstract
{MHD waves can be responsible for plasma fluctuations and short-term
variations of the pulsar emission.
}
{We consider the properties of plane and cylindrical MHD waves that
can exist in the force-free magnetosphere.
} 
{Waves are considered by means of a linear analysis of the force-free
MHD equations. 
} 
{We argue that these particular types of MHD waves can exist in the
magnetosphere of pulsars. These waves are closely related to the Alfv\'{e}n
waves of the standard magnetohydrodynamics modified by the force-free
condition and non-zero charge density. We derive the dispersion equation
fror magnetospheric waves and show that the wave periods are likely within
the range $\sim 10^{-2}-10^{-6}$ s depending on the magnetospheric
parameters.      
}
{}

\authorrunning{V.Urpin: MHD waves in the 
pulsar magnetosphere}

\maketitle

\section{Introduction} 

The magnetospheres of pulsars consist of electron-positron plasma 
with, possibly, some amounts of ions. This plasma can affect the
radiation produced in the inner region of the magnetosphere or at
the stellar surface. Therefore, understanding the properties of a
magnetosphere is of crucial importance for the interpretation of
observations. The growing observational data on spectra and pulse
profiles of isolated pulsars prompt continued improvement of 
theoretical theoretical models of the pulsar magnetosphere (see, 
e.g., Goodwin et al. 2004, Contopoulos et al. 1999, Komissarov 2006). 
Apart from a quasi-static structure, however, various non-stationary 
phenomena in the magnetosphere may affect the radiation. Electromagnetic 
waves in pulsar plasma have been studied extensively over the past few 
decades. The properties of the low-frequency electromagnetic waves
are of central importance for understanding the underlying processes
in the formation of the radio spectrum. Early studies mainly concentrated
on the relativistic plasma flow, assuming a cold or only mildy 
relativistic distribution of electrons and positrons in the plasma rest 
frame (see, e.g., Melrose 1996 and reference therein). Low frequency 
waves were studied by Arons and Barnard (1986), where many of the results 
of the previous studies were rederived and generalized. In all these
cases, the plasma was assumed to be one-dimensional which is justified 
for plasma in a strong magnetic field. The electrostatic oscillations 
with a low frequency have been studied recently by Mofiz et al. (2011),
who found that the thermal and magnetic pressures can generate oscillations
that propagate in the azimuthal direction near the equator.      

Magnetohydrodynamic waves in the pulsar magnetosphere are studied in less
detail. Perhaps the only exception are diocotron modes, which are the 
non-neutral plasma analog of the Kelvin-Helmholtz modes. These modes
have been studied extensively in the context of laboratory plasma devices 
(see, e.g., Levy 1965, Davidson 1990, Davidson \& Felice 1998). However,
diocotron modes can occur also in charged pulsar magnetospheres. The
existence around pulsars of a differentially rotating equatorial disc
with non-vanishing charge density could trigger a shearing instability of
diocotron modes (Petri et al. 2002). In the non-linear regime, the 
diocotron instability can cause diffusion of the charged particles across 
the magnetic field lines outwards (Petri et al. 2003). This turbulent 
charge transport could bear on the problem of electric current closure
in the pulsar inner magnetosphere. The role of a diocotron instability in 
causing drifting sub-pulses in radio pulsar emission has been considered 
by Fung et al. (2006). Note that the diocotron modes should be 
substantially suppressed in a neighbourhood of the light cylinder where 
relativistic effects become important (Petri 2007). The non-axisymmetric
diocotron instability has been observed in 3D numerical modelling of the
pulsar magnetosphere by Spitkovski \& Arons (2002). The properties of 
MHD waves should be very particular in the magnetosphere pulsars since
the electromagnetic energy density in the magnetosphere is greater 
than the kinetic and thermal energy density for typical values of the 
magnetic field. This suggests that over much of the magnetosphere, the 
force-free equation will be a good approximation for various MHD
phenomena, including waves. In this {\it Letter} we consider MHD waves 
that can exist in the pulsar magnetosphere.

\section{Basic equations}

Despite uncertainties in the estimate of many parameters, plasma in the
pulsar magnetosphere is likely collisional and the Coulomb mean free path 
of electrons (and positrons) is shorter than the characteristic length
scale. Therefore, the magnetohydrodynamic description can be justified in 
this plasma. The partial MHD momentum equation for the electrons and
positrons can be obtained in the standard way - multiplying the Boltzmann
kinetic equation by the velocity and integrating over velocity (see, e.g.,
Braginskii 1965). The sum of electron and positron momentum equations 
yields the well-known force-free condition if one neglects terms proportional
electron mass and gas pressure, 
\begin{eqnarray}
\rho_e {\bf E} + \frac{1}{c} \; {\bf j} \times {\bf B} = 0,
\end{eqnarray} 
where $\rho_e = e (n_p - n_e)$ is the charge density. Taking the difference
between electron and positron momentum equations, one obtains Ohm's law
in the magnetosphere
\begin{equation}
{\bf j} = \rho_e {\bf V} + \sigma \!\left({\bf E} \! + \!
\frac{{\bf V}}{c} \! \times \! {\bf B} \right) ,
\end{equation}
where $\sigma = e^2 n_p \tau_e/m_e$ is the conductivity and ${\bf V}$
is the plasma velocity.

Eqs.(1) and (2) can be rewritten as
\begin{eqnarray}
\rho_e {\bf E}^* + \frac{1}{c} \; {\bf J} \times {\bf B} = 0,
\;\;\; {\bf J} - \sigma {\bf E}^* = 0,
\end{eqnarray}
where
\begin{equation}
{\bf E}^* = {\bf E} + \frac{{\bf V}}{c} \! \times \! {\bf B},
\;\;\;  {\bf J} = {\bf j} - \rho_e {\bf V}.
\end{equation}
If $\rho_e \neq 0$, one can exclude ${\bf E}^*$ from Eqs.(3) and obtain 
the equation containing ${\bf J}$ alone,
\begin{equation}
{\bf J} + \frac{\sigma}{c \rho_e} {\bf J} \times {\bf B} = 0.
\end{equation}
It follows immediately from this equation that ${\bf J}=0$. Then, Eqs.(3)
yields ${\bf E}^* = 0$. Hence, we have
\begin{equation}
{\bf j} = \rho_e {\bf V}\;, \;\;\;
{\bf E} =  
- \frac{{\bf V}}{c} \times {\bf B}.
\end{equation}
These equations imply that the force-free condition (1) and Ohm's law
(2) are equivalent to the conditions that the magnetic field is frozen-in
and the only advective current exists in the magnetosphere. Or, in 
other words the force-free condition (1) and Ohm's law (2) are compatible
in a charged plasma only if the electric current is advective and the 
magnetic field is frozen-in. Note that expression (6) can be applied 
only in the regions where $\rho_e \neq 0$.

Eqs.(6) should be complemented by the partial continuity equations for
electrons and positrons
\begin{equation}
\frac{\partial n_{\alpha}}{\partial t} + \nabla \cdot (n_{\alpha} {\bf
V}_{\alpha} ) = \Gamma_{\alpha} = \Gamma^{(+)}_{\alpha} -  
\Gamma^{(-)}_{\alpha},
\end{equation}
where $\Gamma^{(+)}_{\alpha}$ and $\Gamma^{(-)}_{\alpha}$ are the rates
of generation and annihilation of particles of the sort $\alpha$ and
${\bf V}_{\alpha}$ is their partial velocity. Since electrons and 
positrons are generated and annihilate in pairs, we have 
$\Gamma^{(+)}_{e} = \Gamma^{(+)}_{p}$ and $\Gamma^{(-)}_{e} =
\Gamma^{(-)}_{p}$. The difference of the positron and electron
equations yield the charge conservation law,
\begin{equation}
\frac{\partial \rho_e}{\partial t} + \nabla \cdot {\bf j} = 0.
\end{equation} 
The sum of partial equations (7) yields the equation for the total number
density
\begin{equation}
\frac{\partial n}{\partial t} + \nabla \cdot ( n {\bf V}) = 2 \Gamma_e.
\end{equation}

\section{Equation for MHD waves in the pulsar magnetosphere}

MHD processes in the force-free pulsar magnetosphere are governed by
Eqs.(6), (8), and (9) complemented by the Maxwell equations
\begin{eqnarray}
\nabla \cdot {\bf E} = 4 \pi \rho_e , \;\;\; \nabla \times
{\bf E} = - \frac{1}{c} \frac{\partial {\bf B}}{\partial t} ,
\nonumber \\
\nabla \cdot {\bf B} = 0 ,\;\;\; 
\nabla \times {\bf B} = \frac{1}{c} 
\frac{\partial {\bf E}}{\partial t} + \frac{4 \pi}{c} {\bf j}, 
\nonumber \\
{\bf j} = \rho_e {\bf V}, \;\;\;
{\bf E} =  - \frac{{\bf V}}{c} \times {\bf B},
\nonumber \\ 
\frac{\partial n}{\partial t} + \nabla \cdot ( n {\bf V}) = 2 \Gamma_e, 
\;\;\;  \frac{\partial \rho_e}{\partial t} + \nabla \cdot {\bf j} = 0.
\end{eqnarray}  
The force-free MHD phenomena are very particular and this point can be
illustrated by considering linear MHD waves. We assume that the electric
and magnetic fields are equal to ${\bf E}_0$ and ${\bf B}_0$ in the 
unperturbed magnetosphere. The corresponding electric current and
charge density are ${\bf j}_0$ and $\rho_{e0}$, respectively. For the 
sake of simplicity, we neglect hydrodynamic motions in the magnetosphere
and focus on the effects caused by electric currents. Linearizing 
Eqs.(10), we obtain the set of linear equations for waves of a small
amplitude. Small perturbations will be indicated by subscript 1. We
here consider the waves with a short wavelength and space-time 
dependence $\propto \exp(i \omega t - i {\bf k} \cdot {\bf r})$, where
$\omega$ and ${\bf k}$ are the frequency and wave vector, respectively.
These waves exist if their wavelength $\lambda = 2 \pi / k$ is short
compared to the characteristic length scale of the magnetosphere, $L$.
For these perturbations Eqs.(10) takes the form 
\begin{eqnarray}
{\bf k} \cdot {\bf E}_1 = 4 \pi \rho_{e1} , \;\;\; 
c {\bf k} \times {\bf E}_1 =  \omega {\bf B}_1 , \;\;\;
{\bf k} \cdot {\bf B}_1 = 0 , 
\nonumber \\
c {\bf k} \times {\bf B}_1 =  4 i \pi {\bf j}_1
- \omega {\bf E}_1 , \;\;\; {\bf j}_1  = \rho_{e0} {\bf V}_1 ,  
\nonumber \\
c {\bf E}_1 =  - {\bf V}_1 \times {\bf B}_0 , \;\;\;
\omega \rho_{e1} = {\bf k} \cdot {\bf j}_1 ,
\nonumber \\ 
i \omega n_1  - i {\bf k} \cdot ( n_0 {\bf V}_1) = 2 \Gamma_{e1}. 
\end{eqnarray}  
We study only MHD modes with $\omega \ll ck$ because one can split 
electromagnetic and hydrodynamic modes in this case. At $\omega \sim
ck$, a consideration becomes quite cumbersome because electromagnetic 
and hydromagnetic modes are strongly coupled. We focus on understanding 
the nature and qualitative features of magnetospheric waves, rather than
the direct relevance for observational implications, and a consideration 
of the simple particular case $\omega \ll ck$ can help in that. For the
sake of simplicity, we consider waves in the region where $\Gamma_e 
\approx 0$. Substituting the frozen-in condition ${\bf E} = - {\bf V}
\times {\bf B}/c$ into the equation $c \nabla \times {\bf E} = -
\partial {\bf B}/ \partial t$ and linearizing the obtained induction 
equation, we have
\begin{equation}
i \omega {\bf B}_1 = i {\bf B}_0 ({\bf k} \cdot {\bf V}_1) -
i {\bf V}_1 ({\bf k} \cdot {\bf B}_0) - ( {\bf V}_1 \cdot \nabla)
{\bf B}_0 .
\end{equation}   
Eliminating ${\bf j}_1$ from two equations in the second line of Eqs.(11),
we obtain the following expression for ${\bf V}_1$
\begin{equation}
{\bf V}_1 = - \frac{i}{4 \pi \rho_{e0}} ( c {\bf k} \times {\bf B}_1 +
\omega {\bf E}_1).
\end{equation}
Substituting the expression for ${\bf E}_1$ obtained from the linearized
frozen-in condition and neglecting terms terms of the order of $(\omega /
ck)^2$, we have
\begin{equation}
{\bf V}_1  + \frac{i \omega}{4 \pi c \rho_{e0}} {\bf B}_0 \times {\bf V}_1 = 
- \frac{ic}{4 \pi \rho_{e0}} {\bf k} \times {\bf B}_1 .
\end{equation}
The perturbation of the charge density can be calculated from the
equation $\rho_{e1} = \nabla \cdot {\bf E}_1/4 \pi$, then
\begin{equation}
\rho_{e1} = \frac{1}{4 \pi c} [ i {\bf B}_0 \cdot ({\bf k} \times
{\bf V}_1) + {\bf V}_1 \cdot (\nabla \times {\bf B}_0) ].
\end{equation}
Eliminating ${\bf B}_1$ from Eqs.(12) and (14) in favour of ${\bf V}_1$
and neglecting again terms $\sim (\omega / ck)^2$, we obtain the following
equation for ${\bf V}_1$
\begin{equation}
4 \pi \rho_{e0} {\bf V}_1 \!=\! \frac{c}{\omega} {\bf k} \times 
[({\bf V}_1 \cdot \nabla) {\bf B}_0 - i {\bf B}_0 ({\bf k} \cdot {\bf V}_1)
+ i {\bf V}_1 ({\bf k} \cdot {\bf B}_0). 
\end{equation}
The scalar production of this equation and vector ${\bf k}$ yields the 
condition
\begin{equation}
({\bf k} \cdot {\bf V}_1) = 0.
\end{equation} 
This equation implies that the longitudinal waves (with ${\bf k} \cdot
{\bf V}_1 \neq 0$) cannot exist in the force-free magnetosphere. Only
transverse waves with the velocity perpendicular to the wave vector 
(${\bf k} \cdot {\bf V}_1 = 0$) can propagate in this magnetosphere.
We have for transverse waves from Eq.(16)
\begin{equation}
4 \pi \rho_{e0} {\bf V}_1 - i \frac{c}{\omega} ({\bf k} \cdot {\bf B}_0)
{\bf k} \times {\bf V}_1 
= \frac{c}{\omega} {\bf k} \times 
({\bf V}_1 \cdot \nabla) {\bf B}_0 . 
\end{equation}
Eq.(18) is the basic equation governing linear MHD waves in the force-free
magnetosphere. Note that transformations from Eq.(11) to Eq.(18) were
made taking into account terms of the two lowest orders in $\lambda /L$.
The term on the r.h.s. of Eq.(18) is, in general, of the order of  
$\lambda /L$ compared to the second term on the l.h.s. However, this is
not the case if ${\bf k}$ is approximately perpendicular to ${\bf B}_0$
when the term on the r.h.s. becomes dominating.

\section{MHD waves in the pulsar magnetosphere}

{\it Plane waves.} Consider initally the case of plane waves with $({\bf k}
\cdot {\bf B}) \gg \lambda/L$ where we can meglect the term on the r.h.s.
of Eq.(18). We have for these waves
\begin{equation}
4 \pi \rho_{e0} {\bf V}_1 - i \frac{c}{\omega} ({\bf k} \cdot {\bf B}_0)
{\bf k} \times {\bf V}_1 = 0. 
\end{equation}
The dispersion relation corresponding to this equation is
\begin{equation}
\omega^2 = \frac{c^2 k^2 ( {\bf k} \cdot {\bf b})^2 }{\Omega_m^2},
\end{equation}
where $\Omega_m 4 \pi c \rho_{e0}/B_0$ and ${\bf b} = {\bf B}_0/B_0$. It
is convenient to express the characteristic frequency $\Omega_m$ in terms
of the Goldreich-Julian charge density, $\rho_{GJ} = \Omega B_0/2 \pi c$
where $\Omega$ is the angular velocity of a neutron star. Then, we have
$\Omega_m = \xi \Omega$ and $\xi = \rho_{e0}/\rho_{GJ}$, and the dispersion 
equation (20) can be rewritten as
\begin{equation}
\omega = \pm c ( {\bf k} \cdot {\bf b})  \; \frac{ck}{\xi \Omega}.
\end{equation}  
This equation describes the new mode of oscillations that can exist
in the force-free pulsar magnetosphere. Eq.(21) is like the dispersion 
equation for whistlers in ``standard'' plasma. However, there is a principle
difference between the considered waves and whistlers since Eq.(21) contains
the charge density $\rho_{e0} = e (n_{p0} - n_{e0})$, whereas the dispersion
relation for whistlers is determined by $e n_{e0}$. Therefore, the 
magnetospheric waves do not exist if $\rho_{e0} = 0$, but whistlers can
propagate in neutral plasma. The frequency of magnetospheric waves is
higher because it is generally believed that $|n_p - n_e| \ll n_e$ in the
magnetosphere. Deriving Eq.(21), we assumed that $\omega \ll ck$. Therefore,
the considered modes exist if
\begin{equation}
\xi \Omega > c ({\bf k} \cdot {\bf b}).
\end{equation}
This condition can be satisfied for plane waves with the wave vector almost
(but not exactly) perpendicular to the magnetic field, for which the scalar
production $({\bf k} \cdot {\bf b})$ is small. For example, if the angle 
between ${\bf k}$ and ${\bf B}_0$ is $(\pi/2 - \delta)$, $\delta \ll \pi/2$,
then Eq.(22) is satisfied if 
\begin{equation}
\delta < \xi \Omega /ck. 
\end{equation}
Note that generally, magnetospheric waves can exist even if $\delta \sim 1$
but our consideration does not apply to this case.

{\it Cylindrical waves.} Let us assume that the basic magnetic configuration
of a neutron star is dipole and consider a particular sort of waves that can 
exist in a neighbourhood of the magnetic axis. In this neighbourhood, the
field is approximately parallel (or antiparallel) to the axis but the
field component perpendicular to the axis is small. We will mimic the
magnetic geometry near the axis by a cylindrical configuration with the
magnetic field in the $z$-direction. Introducing the cylindrical coordinates
$(s, \varphi, z)$ with the unit vectors $({\bf e}_s, {\bf e}_{\varphi}, 
{\bf e}_z)$, we can model the magnetic field as ${\bf B}_0 = B_0(s) 
{\bf e}_z$. Consider the special case of perturbations with the wavevector
perpendicular to the magnetic field, ${\bf k}=(k_s, k_{\varphi}, 0)$ where
$k_{\varphi}=m/s$ and $m$ is integer. Note that even though we used a
short wavelength approximation deriving Eq.(18), $m$ should not be large 
for cylindrical waves because the cylindrical symmetry of the basic state
is assumed. For these perturbations, Eq.(18) reads
\begin{equation}
4 \pi \rho_{e0} {\bf V}_1 = \frac{c}{\omega} {\bf k} \times {\bf e}_z
V_{1s} \; \frac{d B_0}{d s}.
\end{equation} 
Taking the radial component of this equation, we obtain the dispersion 
relation for cylindrical waves in the form
\begin{equation}
\omega = \frac{c k_{\varphi}}{4 \pi \rho_{e0}} \frac{d B_0}{d s}.
\end{equation}
Cylindrical waves may exist only if $m \neq 0$. These waves propagate 
around the magnetic axis with the velocity $(c/4 \pi \rho_{e0}) (d B_0/d s)$
and period
\begin{equation}
P_m = \frac{2 \pi}{\omega} = \frac{8 \pi^2 s \rho_{e0}}{mc (d B_0/d s)}.
\end{equation}
If we represent the unperturbed charge density as $\xi \rho_{e0}$, then
\begin{equation}
P_m = \frac{2 \xi}{\alpha m} P \left( \frac{s \Omega}{c} \right)^2,
\end{equation}
where $P = 2 \pi/ \Omega$ is the rotation period of a pulsar and $\alpha 
= d \ln B_0 / d \ln s$. The parameter $\alpha$ depends on the magnetic 
configuration. We can estimate it assuming that the poloidal field is
approximately dipole near the axis. The radial and polar components of 
the dipole field in the spherical coordinates $(r, \theta, \varphi)$ are
\begin{equation}
B_r = B_p \left( \frac{a}{r} \right)^3 \cos \theta , \;\;\;
B_{\theta} = \frac{1}{2} B_p \left( \frac{a}{r} \right)^3 \sin \theta ,
\end{equation} 
where $B_p$ is the polar strength of the magnetic field at the neutron
star surface and $a$ is the stellar radius (see, e.g., Urpin et al. 1994).
The radial field component is much greaterthan the polar one near the
magnetic axis and, hence, $\alpha \approx d \ln B_r /d \ln s$. Taking 
into account that $r^2= s^2 + z^2$ and $s \ll z$ in the neighbourhood of
the axis, we obtain with accuracy in terms of the lowest order in $s/r$
that $\alpha = -4 s^2/r^2$. Then, substituting this estimate into Eq.(27),
we have
\begin{equation}
P_m = \frac{\xi}{2 m} P \left( \frac{r \Omega}{c} \right)^2.
\end{equation}
It turns out that the period of cylindrical waves does not depend on the
distance from the magnetic axis if $s \ll r$ but depends on the height above 
the magnetic pole. Therefore, at any given height, perturbations rotate
rigidly around the axis with the period $P_m$. For example, $P_m$ near 
the polar spot at the surface is approximately 
\begin{equation}
P_m \approx 6.3 \times 10^{-6} \; \frac{\xi}{m} P_{0.01}^{-1},
\end{equation}  
where $P_{0.01} = P/ 0.01$s. The period is longer at a higher height,
$r > a$. 

We have considered the cylindrical waves only for the dipole geometry.
Note that in real pulsars, the magnetic field can depart from a simple
dipole geometry and may have a very complex topology even with small-scale 
components (see, e.g., Bonanno et al. 2003, 2005, 2006). The mechanism
of formation of these complex magnetic structures is related to the earliest 
stage of the neutron star life when the star is hydrodynamically unstable.
Dynamo action induced by instabilities generates the magnetic field of 
various length scales that range from the stellar radius to a very short
dissipative scale. These magnetic fields can be frozen into the crust 
that forms when the neutron star cools down. Owing to a high conductivity
of the crust, the magnetic structures can survive a very long time. For
example, structures with the length scale $\geq 1$ km (magnetic spots on
the surface) survive during the entire active life-time of pulsars (Urpin
\& Gil 2004). The complex magnetic topology is inevitable in neutron
stars and can be very important for the magnetospheric structure, 
particularly, in regions close to the star. It is likely, therefore, that
cylindrical magnetic geometries similar to those considered in this section 
can exist in different regions of the magnetosphere.

\section{Discussion}

Our consideration shows that the particular type of waves can exist in
the force-free pulsar magnetosphere. These waves are closely related
to the Alfv\'{e}n waves of the standard magnetohydrodynamics but they
are substantially modified by the force-free condition and non-vanishing
electric charge density in the pulsar magnetosphere. Like Alfv\'{e}n
waves, the magnetospheric ones are transverse because plasma moves only 
in the direction perpendicular to the wavevector, ${\bf k} \cdot {\bf V}_1 =
0$. Our simplified analysis applies only to waves with ${\bf k}$ almost
perpendicular to the magnetic field, ${\bf k} \cdot {\bf B} \ll kB$ but,
in general, these waves can exist for any ${\bf k}$.

The frequency of magnetospheric waves depends on the ratio of the true 
and Goldreich-Julian charge densities, $\xi$. The pair plasma in the
pulsar magnetosphere is likely created in a two-stage process: primary
particles are accelerated by an electric field parallel to the magnetic
field near the poles up to extremely high energy and these produce a
secondary, denser pair plasma via a cascade process (see, e.g., Michel 
1982). The number density of this secondary plasma can exceed the 
Goldreich-Julian number density, $n_{GJ}$, required to maintain a 
corrotation electric field and, hence, the factor $\xi$ can be large. 
Unfortunately, this factor is model-dependent and quite unsertain (see,
e.g., Gedalin et al. 1998, Usov \& Melrose 1995) and can be different
in different regions of the magnetosphere. Therefore, it is difficult to
estimate the periods of waves with a high accuracy but, likely, they
are within a range $10^{-6} - 10^{-2}$ s.

In our analysis, we neglected the destabilizing effects of electric 
currents and shear (that can exist, for example, because of differential 
rotation). These effects may lead to instability of the magnetospheric
waves and exponential growth of small perturbations. Perhaps the instability
caused by electric currents is more efficient in a very strong magnetic 
field. This sort of instability has received considerable attention in the
literature on thermonuclear fusion. The principal features of the pulsar
magnetosphere are the force-free condition and non-zero charge density. 
Nevertheless, in general, the current-driven instability can generate 
the magnetospheric waves. Another mechanism for generating these waves
can be relevant for shear and/or differential rotation. It is well known 
that Alfv\'{e}n waves can be generated in differentially rotating plasma
with the magnetic field by the magnetorotational instability (Velikhov 1959).
This mechanism can operate even if the magnetic field is strong and the 
plasma is magnetized (Urpin \& R\"udiger 2005). Likely, this type of 
mechanism also operate in those regions of the pulsar magnetosphere where
plasma rotates differentially. The generation mechanisms of the magnetospheric
waves will be considered elsewhere. Generation of waves will probably lead 
to fluctuations of the magnetospheric parameters and, hence, the emission
of pulsars. Measurements of the time scale of these fluctuations can
provide information regarding the physical conditions in the magnetosphere.

\end{document}